\documentclass[acmtog, screen, nonacm]{acmart}

\usepackage[dvipsnames]{xcolor}
\usepackage[ruled,vlined]{algorithm2e}

\begin{document}

\title{Neural Image Space Tessellation effect}

\author{Youyang Du}
\affiliation{%
  \institution{Shandong University}
  \city{Jinan}
  \country{China}
}
\affiliation{%
  \institution{Mohamed bin Zayed University of Artificial Intelligence}
  \city{Abu Dhabi}
  \country{United Arab Emirates}
}

\author{Junqiu Zhu}
\affiliation{%
  \institution{Shandong University}
  \city{Jinan}
  \country{China}
}

\author{Zheng Zeng}
\affiliation{%
  \institution{University of California, Santa Barbara}
  \country{USA}
}

\author{Lu Wang}
\affiliation{%
  \institution{Shandong University}
  \city{Jinan}
  \country{China}
}
\authornote{Corresponding authors.}

\author{Lingqi Yan}
\affiliation{%
  \institution{Mohamed bin Zayed University of Artificial Intelligence}
  \city{Abu Dhabi}
  \country{United Arab Emirates}
}
\authornotemark[1]

\renewcommand{\shortauthors}{Du et al.}
\newcommand{\NEW}[1]{\textcolor{Cyan}{#1}}
\newcommand{\TODO}[1]{\textcolor{Red}{#1}}

\settopmatter{printacmref=false} 
\renewcommand\footnotetextcopyrightpermission[1]{} 

\begin{abstract}
    We present Neural Image Space Tessellation effect (NIST), a lightweight screen-space post-processing approach for reducing the faceted silhouettes of low-poly renderings. Instead of tessellating primitives, creating new geometry, or modifying the underlying mesh, NIST uses the low-poly rendering result together with simple auxiliary G-buffer attributes to learn geometry-guided smoothing of object contours in image space. At its core, NIST first deforms image-space contours implicitly and then learns to reassign appearance in the whole image-space, including the deformed regions, preserving texture continuity and avoiding seam artifacts. Experiments show that NIST reduces visually apparent geometric faceting and produces smooth, coherent silhouettes close to tessellation-based smoothing references, with a nearly constant per-frame cost in our tested settings. To the best of our knowledge, NIST is the first work to move the solution of low-poly silhouette faceting from the pre-rendering geometry stage to a post-rendering screen-space stage.
\end{abstract}


\keywords{post-processing, real-time rendering, neural networks}

\begin{teaserfigure}
  \includegraphics[width=\textwidth]{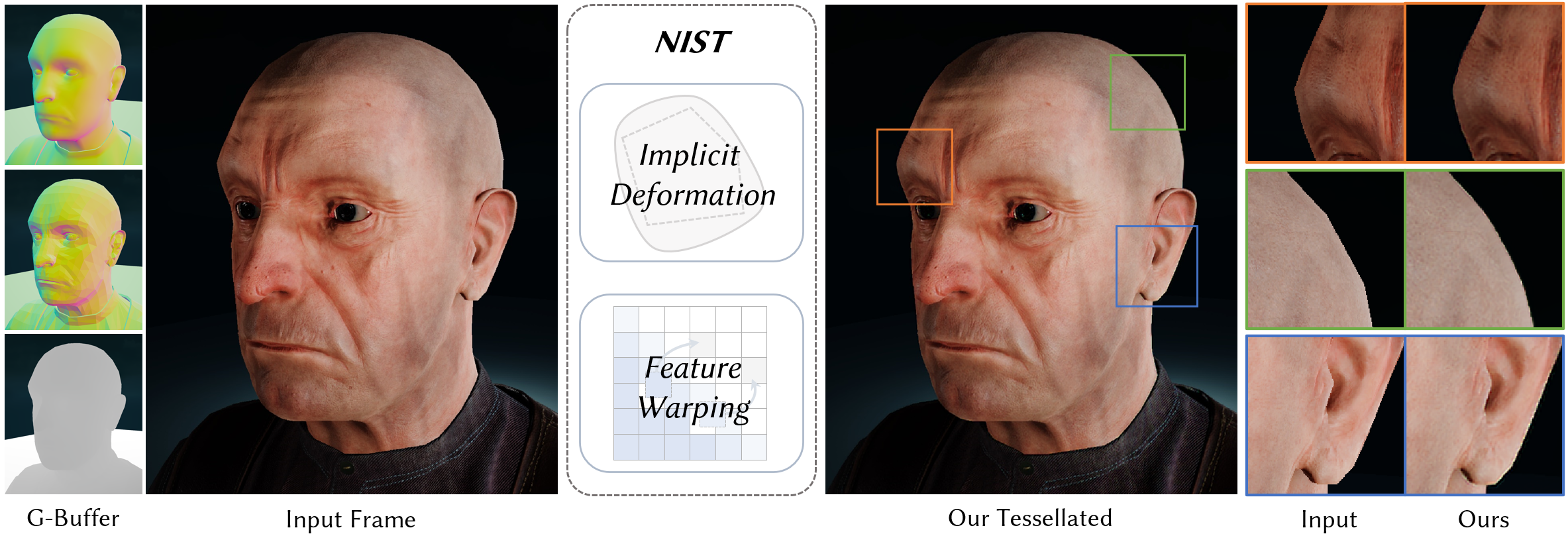}
  \caption{Our NIST method smooths silhouettes directly in screen space via lightweight G-buffer inputs. Without modifying the underlying geometry, our method reduces visually apparent faceting artifacts and produces coherent silhouettes close to tessellation-based smoothing references, as illustrated by the close-up comparisons on the right.}
  \Description{The teaser.}
  \label{fig:teaser}
\end{teaserfigure}

\maketitle

\section{INTRODUCTION}

Low-polygon models remain common in real-time rendering because they are cheap to store, animate, and render. However, when such models are viewed close-up or against a clear background, their contours often reveal visible polygonal faceting. Traditional tessellation and smoothing techniques can reduce these artifacts by increasing geometric resolution and deforming the surface before rasterization \cite{CATMULL1978350, Loop1987SmoothSS, Vlachos2001CurvedPT, Boubekeur2008PhongTessellation}. While effective, these methods add work to the geometry pipeline, and their cost grows with the amount of refined geometry.

In this work, we shift the focus back to the underlying goal: removing the visible geometric faceting that appears in the rendered result. Our goal is not to generate a subdivided mesh, modify the underlying surface, or reproduce a tessellated reference geometrically. Instead, we aim to directly reduce low-poly contour faceting in the rendering result, where the artifact is ultimately observed. This distinction is important: for many real-time applications, a visually smooth contour is sufficient, even if the original geometry remains unchanged.

This motivates a screen-space solution. Contour faceting is view-dependent and is ultimately perceived in the rendered image, which suggests that part of the smoothing effect can be handled after rendering. At the same time, this problem is not a standard image enhancement task. Removing low-poly contour artifacts requires moving image-space boundaries and keeping texture content coherent after the movement. A local filter or a generic image-to-image network can easily blur the boundary or create seams, but it does not directly model the structured deformation needed for contour smoothing.

Therefore, we present \emph{Neural Image Space Tessellation effect} (NIST), a lightweight post-processing method that smooths low-poly contours directly in screen space. Given a low-poly rendering and a small set of standard G-buffer attributes, NIST learns where object contours should be adjusted and how appearance should be reassigned after the refinement. The network first infers implicit image-space contour deformation, and then uses feature warping to propagate color and texture information into the deformed regions. This design separates the structural problem of smoothing the contour from the appearance problem of preserving texture continuity, allowing NIST to produce smooth contours without visible seam artifacts and without changing the input geometry.


To sum up, our contributions are summarized as follows:
\begin{itemize}
    \item To the best of our knowledge, this is the first work to formulate low-poly contour faceting as a screen-space post-processing problem, targeting the visual smoothing effect normally achieved through pre-rendering geometry refinement without tessellating or modifying the mesh.
    \item We propose a geometry-guided neural architecture that combines implicit image-space contour deformation with feature warping, enabling structured contour smoothing while preserving texture continuity.
    \item We evaluate NIST against geometry-based smoothing references and ablated variants, showing that the proposed design reduces visible contour faceting with a small and nearly constant per-frame overhead in our tested settings.
\end{itemize}

\section{RELATED WORK}

\subsection{Traditional Geometric Subdivision and Smoothing}

Geometric subdivision and mesh upsampling methods improve low-poly assets before rasterization by refining object-space geometry. They can be broadly categorized into topology-based subdivision surfaces and local geometric approximations designed for real-time rendering.

\paragraph{Subdivision Surfaces}
Classical subdivision schemes, such as Catmull-Clark \cite{CATMULL1978350} and Loop subdivision \cite{Loop1987SmoothSS}, recursively refine mesh connectivity and reposition vertices to approach smooth limit surfaces. Subsequent work adapted this object-space formulation to real-time and production settings, including approximate GPU-friendly evaluation \cite{Loop2008ApproximatingCS}, creased surface handling \cite{kovacsRealtimeCreased}, and feature-adaptive refinement for high-quality surface evaluation \cite{niesnerFeatureCC}. Neural Subdivision \cite{Liu2020NeuralSubdivision} follows the same object-space coarse-to-fine formulation, but replaces fixed linear vertex averaging with a neural network that predicts new vertex positions from local patch geometry. Although this learned rule can better preserve geometric details, its input remains a coarse mesh and its output is still a denser surface mesh used by the standard rendering pipeline.

\paragraph{Local Geometric Approximation}
Real-time geometric approximation methods pursue a lighter object-space alternative. PN-Triangles \cite{Vlachos2001CurvedPT} construct local curved patches from vertex normals, while PN-AEN \cite{McDonald2010PNAEN} improves watertightness using adjacent-edge information. Phong Tessellation \cite{Boubekeur2008PhongTessellation} further simplifies this idea by projecting triangle vertices according to vertex-normal planes, and was later generalized to polygonal meshes \cite{Hettinga2017PhongTA}. These methods are efficient, but they still operate before rasterization and increase or modify the rendered geometry.

In contrast, NIST targets the visible consequence of insufficient geometric refinement after rasterization. It does not output a refined mesh, add primitives, or modify object-space geometry. Instead, it uses the low-poly rendered frame and lightweight G-buffer cues to infer a small screen-space contour refinement, approximating the visible smoothing effect of geometry-based methods without performing geometric subdivision.

\subsection{Image Warping and Deformation-Driven Synthesis}

Classical image warping and morphing resample image content under prescribed spatial transformations, often using mesh, feature-line, or control-point correspondences to guide deformation \cite{Wolberg1990DigitalImageWarping,Beier1992FeatureMetamorphosis}. Neural warping makes this operation trainable by sampling pixels or features from a source image under a predicted correspondence field. \citet{Jaderberg2015SpatialTransformer} introduced Spatial Transformer Networks (STN), which combine a localization network, a grid generator, and a bilinear sampler into an end-to-end trainable module. This work established the basic neural warping primitive used by many later methods. However, the original STN formulation relies on a predefined transformation family, such as affine, projective, or thin-plate spline transformations. It is therefore well suited to global alignment or smooth low-dimensional deformation, but it is not by itself a general solution for arbitrary, content-adaptive local warping.

Subsequent work increased the flexibility of neural warping by introducing task-specific deformation models. \citet{Ganin2016DeepWarp} predicted dense displacement fields in a coarse-to-fine manner for photorealistic gaze manipulation, showing that direct warping can preserve high-frequency details better than bottlenecked encoder-decoder synthesis. In a different direction, \citet{Cole2017NormalizedFaces} used predicted facial landmarks and differentiable spline interpolation for face normalization, and \citet{Shi2019WarpGAN} predicted sparse control points followed by thin-plate spline warping for caricature generation. These methods illustrate an important trade-off: dense fields are flexible but difficult to constrain, while sparse control-point warping is more interpretable and stable but limited by the chosen control structure.

For larger semantic or articulated motion, later methods often make the target configuration explicit before warping. \citet{Siarohin2018DeformableGAN} and \citet{Ren2020DeepSpatialTransformation} use target poses to guide person-image deformation, while \citet{Siarohin2019AnimatingObjects} and \citet{Siarohin2019FirstOrderMotion} derive motion from a driving image or video through self-supervised keypoints and local motion models. Geometry-guided face-animation methods further use stronger priors, such as 3D face guidance and progressive feature warping \cite{Zhong2022GeometryDrivenPW}. These works show that feature warping is effective for preserving appearance when a target pose, driving motion, or geometric model defines where content should move. NIST adopts this appearance-preserving principle in a different rendering setting: \textit{no target pose or driving frame is provided}, and the desired deformation is only a small screen-space contour refinement inferred from the low-poly rendering and lightweight G-buffer cues. Moreover, our network needs to be integrated into the real-time rendering pipeline, making our target a time-sensitive task that totally differs from the task above.

\subsection{Neural Postprocessing}
\paragraph{Denoising and Supersampling}
Neural post-processing has been widely used in rendering pipelines to reconstruct missing or noisy radiance samples. Monte Carlo denoising methods use rendered color together with auxiliary buffers to recover clean images from sparse samples \cite{Kalantari2015AML,Bako2017KernelpredictingCN,Chaitanya2017InteractiveRO,Vogels2018DenoisingWK,Gharbi2019SamplebasedMC,Yu2021MonteCD}, while neural supersampling, frame interpolation, and frame extrapolation reconstruct spatial or temporal radiance samples for real-time rendering \cite{Xiao2020NeuralSF,Zhong2023FuseSRSR,Briedis2021NeuralFI,Briedis2023KernelBasedFI,Guo2021ExtraNet,Wu2024GFFEGF,Wu2025MoFlowMF}. 

\paragraph{Generative Neural Post-processing}
Recent work has explored diffusion models for rendering post-processing, enabling flexible re-rendering, appearance editing, and relighting by synthesizing images from rendered inputs or intermediate representations \cite{Zeng2024RGBXID,Zeng2025NeuralRemasterPD,Liang2025DiffusionRendererNI}. 

These methods may rely on G-buffers or motion cues, but their primary objective remains radiance reconstruction. In contrast, NIST targets a geometric signal reconstruction problem: it aims to recover the image-space appearance of a smooth surface from a discretized low-poly geometric representation, which requires moving contour structure and reassigning appearance features rather than only reconstructing missing color samples. Although generative models offer strong synthesis capability, we do not adopt a generative formulation because current generative architectures and iterative sampling processes are difficult to optimize to the latency required by real-time rendering.

\section{METHOD}

\subsection{Overview}

\begin{figure}
    \centering
    \includegraphics[width=\linewidth]{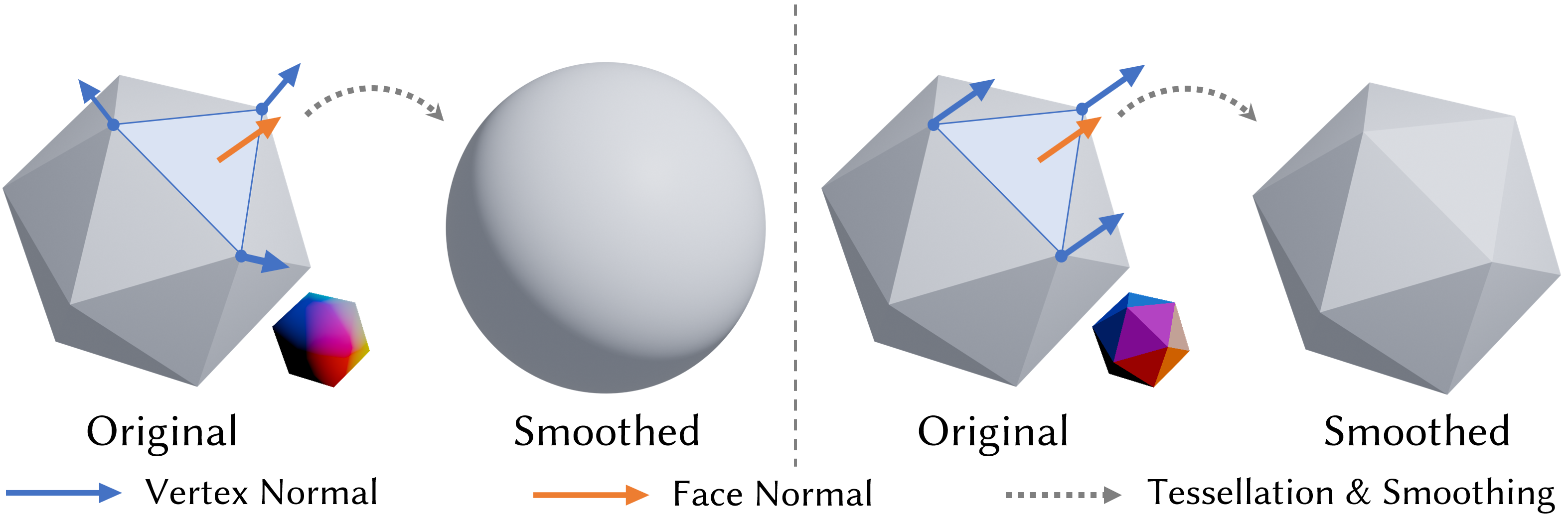}
    \caption{Normal-based guidance for screen-space contour refinement. When shading normals deviate from geometric normals, the rendered appearance suggests a smoother contour than the underlying low-poly geometry provides; when the two normals are consistent, unnecessary deformation should be avoided.}
    \label{fig:guidance-insight}
\end{figure}

Our goal is to reduce the visually apparent faceting of low-poly object contours directly in screen space, without actually modifying the underlying geometry. Specifically, given a low-poly rendered image $I^{input}$ and a set of lightweight G-buffer attributes $G$, NIST predicts a refined image $I^{pred}$ whose visible contours appear smoother while preserving the appearance of the original rendering. In our implementation, $G=\{D,N_g,N_s\}$, where $D$ denotes the depth buffer, $N_g$ denotes geometric normals, and $N_s$ denotes shading normals. We choose these attributes because depth provides screen-space boundary and visibility context, while the discrepancy between $N_g$ and $N_s$ indicates where the low-poly geometry disagrees with the intended smooth appearance. As illustrated in Figure~\ref{fig:guidance-insight}, such normal discrepancy provides a controllable cue for where contour deformation is needed, while normal consistency suggests that deformation should be suppressed.

Prior neural warping, deformation-driven synthesis, and generative rendering post-processing methods suggest possible ways to move or synthesize image content \cite{Ganin2016DeepWarp, Siarohin2019FirstOrderMotion, Liang2025DiffusionRendererNI}. However, NIST targets a time-sensitive real-time rendering setting, where heavy generative models, iterative sampling, and complex semantic motion architectures are difficult to deploy. We therefore adopt a lightweight CNN-based design, but this screen-space formulation still introduces two key difficulties. First, contour refinement requires structured image-space deformation rather than local color filtering, because the visible boundary itself must move in a geometry-aware manner. Second, once the contour has moved, the refined image must assign coherent colors across the entire image space, including regions that become newly visible after the contour refinement. We address these difficulties with a multi-scale \emph{Neural Deformation} algorithm. Each Neural Deformation step contains two complementary modules: an \emph{Implicit Deformation Module}, which estimates contour-aware image-space deformation, and a \emph{Feature Warping Module}, which reassigns appearance features across the image space. In the following sections, we first introduce the core idea of Neural Deformation as the main refinement algorithm, and then detail the design of the Implicit Deformation Module and the Feature Warping Module.

\begin{figure*}
    \centering
    \includegraphics[width=\textwidth]{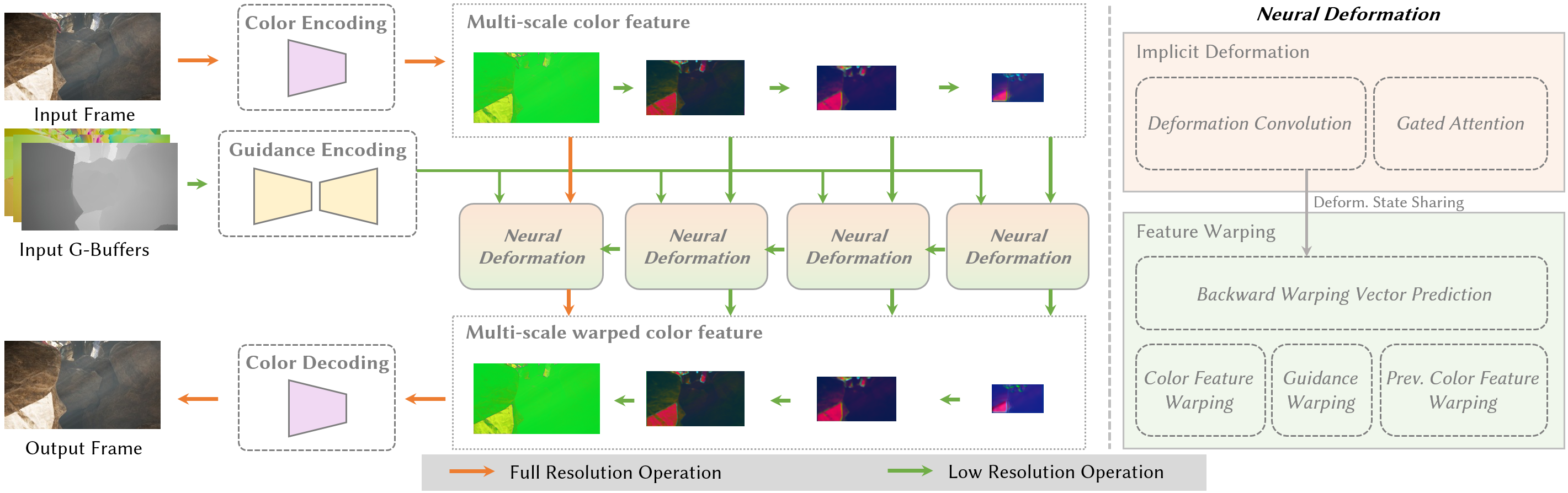}
    \caption{Overview of the NIST pipeline. Given an input frame and G-buffer attributes, the network extracts color and geometry guidance features, applies multi-scale Neural Deformation to update image-space contours and reassign color features, and decodes the final refined image. Orange arrows indicate full-resolution operations, while green arrows indicate reduced-resolution operations.}
    \label{fig:pipeline}
\end{figure*}

\subsection{Neural Deformation}

Neural Deformation is the basic refinement operation used by NIST to change image-space contours and update the corresponding appearance features. The first requirement of this operation is controllable contour deformation. To represent this process, we maintain a \emph{deformation state} $z_d$, which is a latent image, and stores contour and deformation information in image space. Rather than treating deformation as an explicit mesh deformation or a hand-designed contour curve, $z_d$ provides a latent representation from which the network can infer how the current image-space structure should change. Each deformation step takes $z_d$ together with a spatially aligned guidance feature $z_g$, and produces an updated deformation state $z_d^+$ that represents the contour after one deformation update.

The guidance feature $z_g$ is computed entirely from G-buffer attributes and is independent of the rendered color. This design separates the geometric decision of where and how contours should move from the appearance information carried by the color image. After the deformation state has been updated, the method must determine how color features should be distributed in the new image-space configuration. For this purpose, Neural Deformation uses a Feature Warping Module. Given the updated deformation state $z_d^+$ and a color feature map $z_c$, the module predicts a backward warping vector $v$ from $z_d^+$ directly and resamples $z_c$ to obtain the deformed color feature $z_c^+$. The resulting feature has contours aligned with the updated deformation state and carries color information reassigned according to the predicted image-space deformation.

\paragraph{Implicit Deformation Module}
The Implicit Deformation Module updates the deformation-state latent image under geometry guidance. Its inputs are the current deformation state $z_d$ and the aligned guidance feature $z_g$. The module has two core designs. First, it uses a spatial attention mechanism to decide where the deformation state should be updated. The guidance feature provides query-like information derived from G-buffers, while the deformation state provides the latent contour structure to be transformed:
\begin{equation}
    Q = F_q(z_g), \quad K = F_k(z_d), \quad V = F_v(z_d),
\end{equation}
\begin{equation}
    W_d = \sigma\left(F_a([Q,K])\right), \quad \tilde{z}_d = W_d \odot V .
\end{equation}
Here, $F_q$, $F_k$, $F_v$, and $F_a$ are learnable convolutions, $[\cdot,\cdot]$ denotes channel-wise concatenation, $\sigma(\cdot)$ denotes the activation function that maps the predicted spatial weights to $[0,1]$, and $\odot$ denotes element-wise multiplication. The result $\tilde{z}_d$ is a reweighted deformation state, where geometry-derived weights control which image-space locations should contribute to the deformation update.

Second, the module applies a deformation convolution block with an enlarged receptive field to the reweighted deformation state. This block allows each contour location to aggregate a wider neighborhood before updating the latent structure, which is important because contour smoothing is not a purely per-pixel operation. The updated deformation state is
\begin{equation}
    z_d^+ = \Phi_d(\tilde{z}_d),
\end{equation}
where $\Phi_d$ denotes the deformation convolution block. The output $z_d^+$ is still a latent image rather than an explicit deformation field. It encodes the deformed contour structure that will be used by the subsequent warping step.

\paragraph{Feature Warping Module}
The Feature Warping Module converts the updated deformation state into appearance reassignment. Given $z_d^+$, the module predicts a backward warping vector
\begin{equation}
    v = \Phi_w(z_d^+),
\end{equation}
where $\Phi_w$ denotes the warping-vector predictor. The vector $v$ defines, for each target location in the refined image space, where the corresponding color feature should be sampled from the previous image-space feature map. We then obtain the deformed color feature by
\begin{equation}
    z_c^+ = \mathrm{Warp}(z_c, v),
\end{equation}
where $\mathrm{Warp}(\cdot)$ is implemented as differentiable backward warping. This backward formulation avoids the hole problem usually caused by forward warping and produces a dense feature map in the deformed image space. As a result, the network can move contours while preserving texture continuity, because appearance is reassigned through feature sampling rather than synthesized from scratch.


\subsection{Multi-scale Neural Deformation}

A single Neural Deformation step updates contours at one feature resolution. However, low-poly contour artifacts contain both large-scale contour errors and fine-scale boundary misalignments. Predicting all deformation at full resolution is inefficient and tends to rely on local evidence, while operating only at a coarse resolution loses the detail needed for accurate boundary and texture alignment. We therefore apply Neural Deformation in a coarse-to-fine manner, so that low-resolution steps establish large-scale coherent contour deformation and high-resolution steps refine local boundaries and appearance continuity.

Let $\{z_c^{(t)}\}_{t=1}^{T}$ denote the color feature pyramid extracted from the input image, ordered from coarse to fine. The G-buffer encoder produces geometry guidance features, which are resized to the corresponding scale and denoted as $z_g^{(t)}$. Since no previous deformation state exists at the coarsest scale, we initialize it from the coarsest guidance feature using learnable convolutions:
\begin{equation}
    z_d^{(1)} = \Phi_{init}\left(z_g^{(1)}\right).
\end{equation}
For subsequent scales, the deformation state is propagated from the previous coarser scale. At each scale $t$, the color information comes from two sources: the encoder feature $z_c^{(t)}$, which is still aligned with the undeformed input image, and the upsampled deformed color feature propagated from the previous coarser scale, denoted as $p_c^{(t)}$. The propagated feature is omitted at the coarsest scale. The network then applies one Neural Deformation step:
\begin{equation}
    z_d^{(t)+}, z_c^{(t)+}, v^{(t)} =
    \mathrm{ND}\left(z_d^{(t)}, z_g^{(t)}, z_c^{(t)}, p_c^{(t)}\right),
\end{equation}
where $z_d^{(t)+}$ is the updated deformation state, $z_c^{(t)+}$ is the color feature after appearance reassignment, and $v^{(t)}$ is the backward warping vector predicted at the current scale. The updated deformation state is propagated to the next finer scale, providing a structural prior for subsequent contour refinement.

Because different scales are processed sequentially, their feature maps do not originally live in the same image-space configuration. Features extracted directly from the encoder remain aligned with the undeformed input image, while features propagated from coarser scales have already been warped by previous Neural Deformation steps. To keep them consistent, we maintain a cumulative backward warping vector $\tilde{v}^{(t)}$:
\begin{equation}
    \tilde{v}^{(t)} =
    v^{(t)} + \mathrm{Up}\left(\tilde{v}^{(t-1)}\right),
\end{equation}
where $\mathrm{Up}(\cdot)$ resizes the previous cumulative vector to the current scale. This cumulative vector records the deformation history from coarse scales to the current scale, and is used to align newly extracted color features with the current deformed image-space configuration.

Concretely, the current encoder feature is warped by the cumulative vector, while the propagated deformed feature, when available, is further aligned by the current-scale vector:
\begin{equation}
    \bar{z}_c^{(t)} = \mathrm{Warp}\left(z_c^{(t)}, \tilde{v}^{(t)}\right), \quad
    \bar{p}_c^{(t)} = \mathrm{Warp}\left(p_c^{(t)}, v^{(t)}\right), \quad t>1 .
\end{equation}
The aligned features are fused by a learnable convolution to form $z_c^{(t)+}$, which is then passed to the next finer scale as $p_c^{(t+1)}$; at the coarsest scale, the fusion uses only the warped encoder feature.

This coarse-to-fine design gives different scales complementary roles. Coarse levels have larger effective receptive fields and therefore capture low-frequency, coherent contour deformation. Finer levels operate after the main contour structure has been established, allowing the network to correct local boundary errors and preserve high-frequency appearance details. As a result, Multi-scale Neural Deformation performs structured contour refinement while keeping color features consistently reassigned across the entire image space. After the finest scale is processed, the final deformed color feature is decoded into the output image:
\begin{equation}
    I^{pred} = \Phi_{dec}\left(z_c^{(T)+}\right),
\end{equation}
where $\Phi_{dec}$ denotes the color decoder.

\subsection{Loss Function}

During training, we supervise NIST with a reference image $I^{ref}$ rendered from the same scene using Phong Tessellation \cite{Boubekeur2008PhongTessellation} under matched camera, material, lighting, and render-target settings. This reference is used only as an image-space supervision target; it is not required during inference and does not imply that NIST reconstructs tessellated geometry. The predicted image $I^{pred}$ is optimized with a composite objective that emphasizes deformation regions while preserving perceptual appearance.

The primary term is a residual-relative reconstruction loss. Since most pixels remain similar between the low-poly input and the reference, directly averaging pixel errors can underweight the regions where contour refinement is needed. We therefore normalize the prediction error by the reference-input residual:
\begin{equation}
    \mathcal{L}_{RR} =
    \frac{1}{N}\sum_i
    \frac{\left\|I_i^{pred}-I_i^{ref}\right\|_1}
    {\left\|I_i^{ref}-I_i^{input}\right\|_1+\epsilon},
\end{equation}
where $N$ is the number of pixels and $\epsilon=10^{-6}$ avoids division by zero. This loss gives larger relative importance to pixels where the reference differs from the low-poly input, which typically occur near refined contours.

To further emphasize visible contour misalignment, we use a top-$k$ residual loss. Let
\begin{equation}
    e_i=\left\|I_i^{pred}-I_i^{ref}\right\|_1
\end{equation}
be the per-pixel prediction residual, and let $S_k$ denote the indices of the $k$ largest residuals in the image. The loss is
\begin{equation}
    \mathcal{L}_{top}=\frac{1}{k}\sum_{i\in S_k} e_i .
\end{equation}
This term focuses optimization on the most visible errors, which are often caused by inaccurate contour deformation or appearance reassignment near boundaries.

Finally, we include the LPIPS perceptual loss \cite{zhang2018unreasonableeffectivenessdeepfeatures} to discourage overly smooth outputs and preserve high-frequency appearance details. The full objective is
\begin{equation}
    \mathcal{L} =
    \lambda_{RR}\mathcal{L}_{RR}
    + \lambda_{top}\mathcal{L}_{top}
    + \lambda_{LPIPS}\mathcal{L}_{LPIPS}.
\end{equation}
In all experiments, we set $\lambda_{RR}=0.1$, $\lambda_{top}=10$, and $\lambda_{LPIPS}=30$.

\subsection{Implementation Details}

We transform both geometric normals and shading normals into camera space before feeding them into the network, which makes the guidance representation consistent across viewpoints. In our implementation, geometric normals correspond to face normals, and shading normals correspond to interpolated vertex normals; we do not use normal maps as additional input. The input G-buffer is encoded by a geometry guidance encoder, while the rendered color image is encoded into a multi-scale color feature pyramid.

Although the input and output images share large overlapping regions, NIST predicts the final refined image directly rather than predicting a residual image. This choice follows from the feature warping formulation: Neural Deformation reassigns color features across the whole image space, including newly revealed regions introduced by contour deformation, rather than only applying localized color corrections to the input image.

To reduce computation, the deformation and feature warping operations are performed at a reduced internal resolution in our reported implementation, while the first-level color feature processing is kept at full resolution to preserve fine image details. Unless otherwise specified, learnable convolutions use a double-convolution block consisting of two $3\times3$ convolutions with LeakyReLU activations. The deformation convolution block uses a $7\times7$ first convolution followed by a $3\times3$ convolution, providing a wider receptive field for contour updates.

The network is implemented in PyTorch and optimized using Adam with a learning rate of $1\times10^{-4}$ and a weight decay of $1\times10^{-5}$. All models are trained on a single NVIDIA A100 GPU with a batch size of 4 for approximately 24 hours.

\newcommand{\SoulCave}{{\textsc{SoulCave}}}

\section{RESULTS AND COMPARISONS}

\begin{table}[t]
\centering
\caption{Dataset statistics for the evaluated scenes. We report the number of training and testing frames, as well as the maximum number of visible mesh faces per frame.}
\label{tab:dataset_stats}
\begin{tabular}{l | c c c}
\hline
Scene & Training Set & Testing Set & Max. Visible Faces \\
\hline
\textsc{Junkyard} & 2000 & 500  & 39,621K \\
\textsc{SoulCave} & 4000 & 540  & 584.4K  \\
\textsc{Cowboy}   & 2500 & 500  & 462.2K  \\
\textsc{Bronze}   & 3000 & 500  & 780.4K  \\
\hline
\end{tabular}
\end{table}

To evaluate the effectiveness of our NIST method, we compare our method against a representative traditional tessellation approach. Specifically, we use the PN-Triangles \cite{Vlachos2001CurvedPT} algorithm as implemented in Unreal Engine 4.27 as our reference. Our evaluation is conducted on four scenes, each split into disjoint training and testing sets, as shown in Table \ref{tab:dataset_stats}. In addition to comparisons with traditional methods, we perform a series of ablation studies to analyze the key components of our framework. All experiments are executed on a desktop PC equipped with an Intel i9-14900K CPU and an NVIDIA RTX 5090 GPU.

\subsection{Qualitative and Quantitative Comparison}
\begin{figure*}
    \centering
    \includegraphics[width=0.99\linewidth]{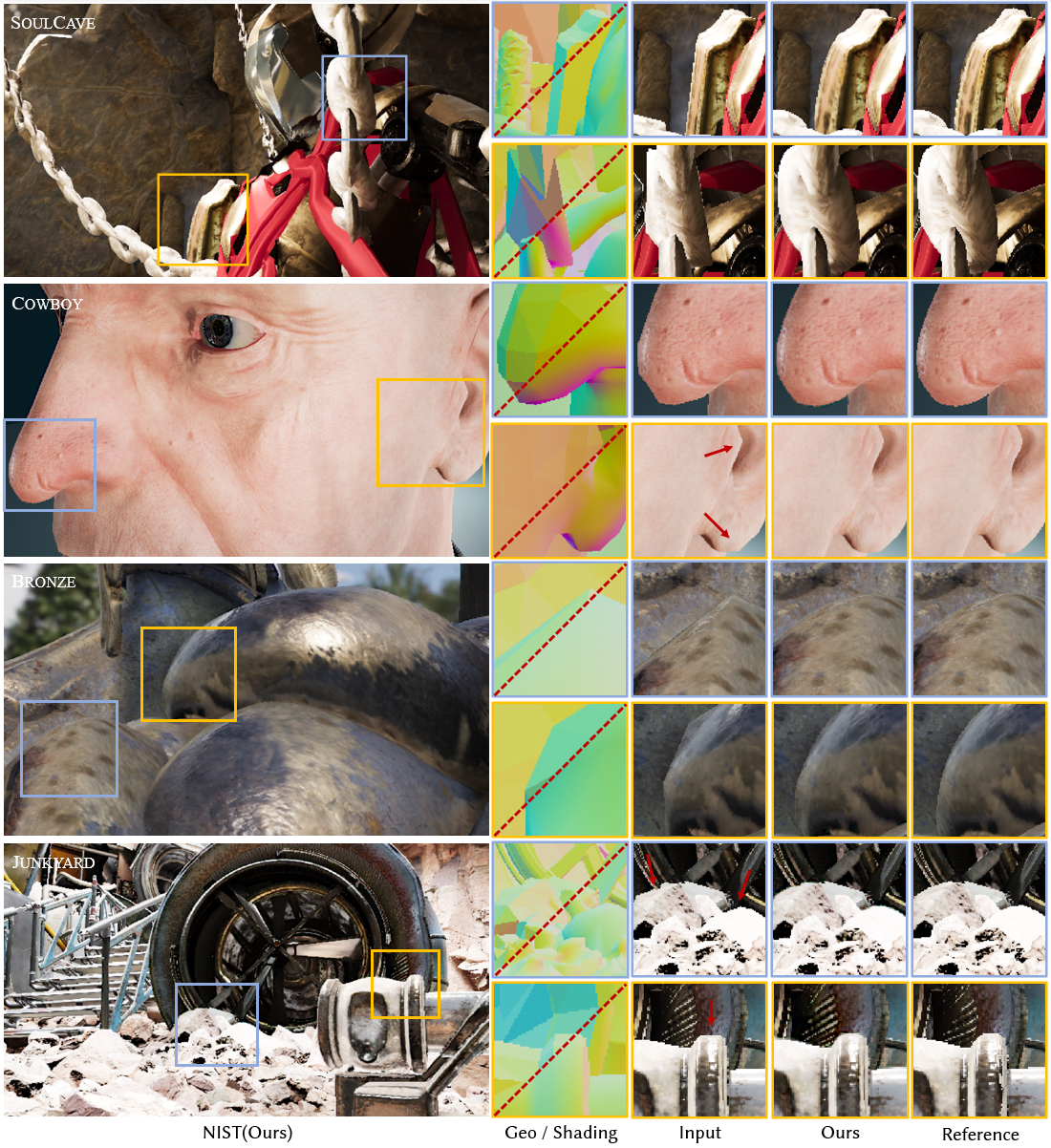}
    \caption{Qualitative comparison across four scenes. From top to bottom: \textsc{SoulCave}, \textsc{Cowboy}, \textsc{Bronze}, and \textsc{Junkyard}. The left column shows the output of NIST. For each scene, two representative regions are highlighted and magnified on the right, comparing the input rendering, our result, and the reference. We also show the geometric normals and shading normals in the middle. NIST effectively removes silhouette artifacts while producing results visually comparable to geometric tessellation.}
    \label{fig: quality comparison}
\end{figure*}

Figure \ref{fig: quality comparison} presents qualitative comparisons between our method and traditional tessellation across four representative scenes. For each scene, we show the original input rendering without any contour smoothing, the result produced by PN-Triangles as implemented in Unreal Engine 4.27, and the output of NIST. In addition, two representative regions are highlighted and magnified for each frame to facilitate close inspection of contour quality and local surface continuity.

Across all scenes, NIST effectively removes visually distracting contour artifacts present in the unsmoothed input, producing results that visually resemble the contour smoothing obtained with geometric tessellation. In scenes such as \SoulCave, which exhibit large-scale geometric deformation, NIST successfully performs substantial image-space contour refinement, recovering smooth and coherent outlines without introducing visible distortions. In contrast, scenes featuring organic surfaces, such as \textsc{Cowboy}, benefit from NIST’s ability to perform subtle, fine-grained smoothing, resulting in visually consistent contours that preserve the original appearance. The zoomed-in regions further demonstrate that NIST maintains high-frequency appearance details and produces seamless transitions along refined contours, despite operating entirely in screen space and leaving the geometry unchanged.

\begin{figure*}
    \centering
    \includegraphics[width=\linewidth]{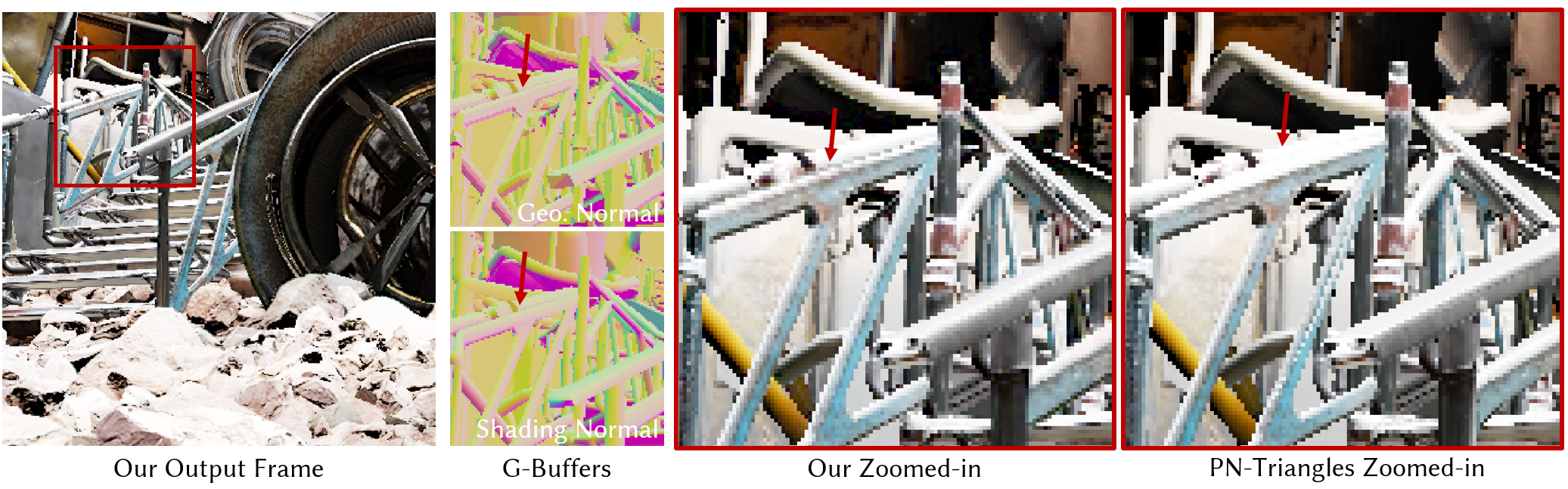}
    \caption{Preservation of straight element guided by normal consistency. In regions where geometric normals and shading normals agree (middle), NIST correctly avoids deformation and preserves straight structures (left, zoom-in), while PN-Triangles introduces undesired deformation (right, zoom-in).}
    \label{fig: sharpness}
\end{figure*}

We also observe that NIST is able to avoid unnecessary smoothing in regions where geometric normals and shading normals are consistent. Figure \ref{fig: sharpness} shows an additional example from the \textsc{Junkyard} scene, where straight structural elements are preserved by our method, while PN-Triangles introduces undesired deformation despite the absence of perceptual inconsistency. This behavior highlights the effectiveness of our normal-based guidance in selectively applying deformation only where refinement is required.

\subsection{Temporal Stability}
To evaluate temporal stability, we compute motion-compensated luminance Flicker on consecutive frames. For a sequence $X_t$ and motion vectors $F_t$, we warp the previous-frame luminance to the current frame and measure the mean absolute residual over valid warped pixels:
\begin{equation}
    \mathcal{F}(X;F)=\frac{1}{T-1}\sum_{t=2}^{T}
    \frac{\sum_x M_t(x)\left|Y(X_t)(x)-W(Y(X_{t-1}),F_t)(x)\right|}
    {\sum_x M_t(x)+\epsilon},
\end{equation}
where $$Y(X_t) = 0.2126 X_{t,R} + 0.7152 X_{t,G} + 0.0722 X_{t,B}$$ converts RGB to luminance, $W(\cdot)$ denotes motion-vector warping, and $M_t$ is the valid-warp mask. We use only this valid mask in our evaluation to exclude samples whose warped coordinates fall outside the image domain. The valid ratio reported in Table \ref{tab:temporal_flicker} is the average fraction of pixels retained by this mask, i.e., $\sum_x M_t(x)/(HW)$, and indicates the effective image area used for the Flicker measurement. During evaluation, input flicker is computed using the motion vectors from the low-poly rendering, while both NIST output flicker and reference flicker are computed using the motion vectors from the rendering of the scene with Phong Tessellation.

We evaluate temporal stability on \SoulCave, \textsc{Cowboy}, and \textsc{Bronze}, the three scenes where image-space deformation is visually pronounced and therefore temporally measurable. The low-poly input is evaluated using its own motion vectors, while the reference and NIST outputs are evaluated using the reference motion vectors, since NIST aims to match the smooth reference appearance. As shown in Table \ref{tab:temporal_flicker}, the Flicker of NIST remains close to the reference across all three scenes. This indicates that the proposed screen-space deformation does not introduce substantial additional frame-to-frame instability under the target smooth motion field.

\begin{table}[t]
\centering
\caption{Motion-compensated temporal Flicker on scenes with pronounced image-space deformation. We report mean luminance Flicker under a valid-warp mask; lower values indicate better temporal stability.}
\label{tab:temporal_flicker}
\begin{tabular}{l|ccc|c}
\hline
Scene & \multicolumn{3}{c|}{Flicker ($\times 10^{-2}$)} & Valid Ratio \\
\cline{2-4}
      & Input & Reference & NIST & \\
\hline
\SoulCave        & 1.201 & 1.191 & 1.193 & 96.83\% \\
\textsc{Cowboy} & 2.275 & 2.268 & 2.238 & 98.28\% \\
\textsc{Bronze} & 1.177 & 1.256 & 1.245 & 93.49\% \\
\hline
\end{tabular}
\end{table}

\begin{table}[t]
\centering
\caption{Inference latency breakdown (mean, in ms) at different input resolutions. H2D denotes Host-to-Device transfer, and D2H denotes Device-to-Host transfer.}
\label{tab:performance}
\begin{tabular}{c|ccc|c}
\hline
Resolution  & H2D & \textbf{GPU Compute} & D2H & \textbf{Total} \\
\hline
360p  & 0.145 & \textbf{4.520} & 0.034 & \textbf{4.699} \\
720p  & 0.537 & \textbf{5.000} & 0.136 & \textbf{5.673} \\
1080p & 1.190 & \textbf{6.186} & 0.304 & \textbf{7.680} \\
\hline
\end{tabular}
\end{table}

\subsection{Resolution Scaling and Performance}
To evaluate the scalability of NIST with respect to image resolution, we report its inference performance under different rendering resolutions. Table \ref{tab:performance} summarizes the inference latency measured at 360p, 720p, and 1080p in a representative scene. As shown in the table, the inference cost of NIST increases modestly with image resolution instead of a linear increment. Importantly, this cost is independent of scene geometric complexity, in contrast to geometry-based tessellation methods whose performance scales with mesh complexity.

\subsection{Ablation Study}
\begin{figure*}
    \centering
    \includegraphics[width=\linewidth]{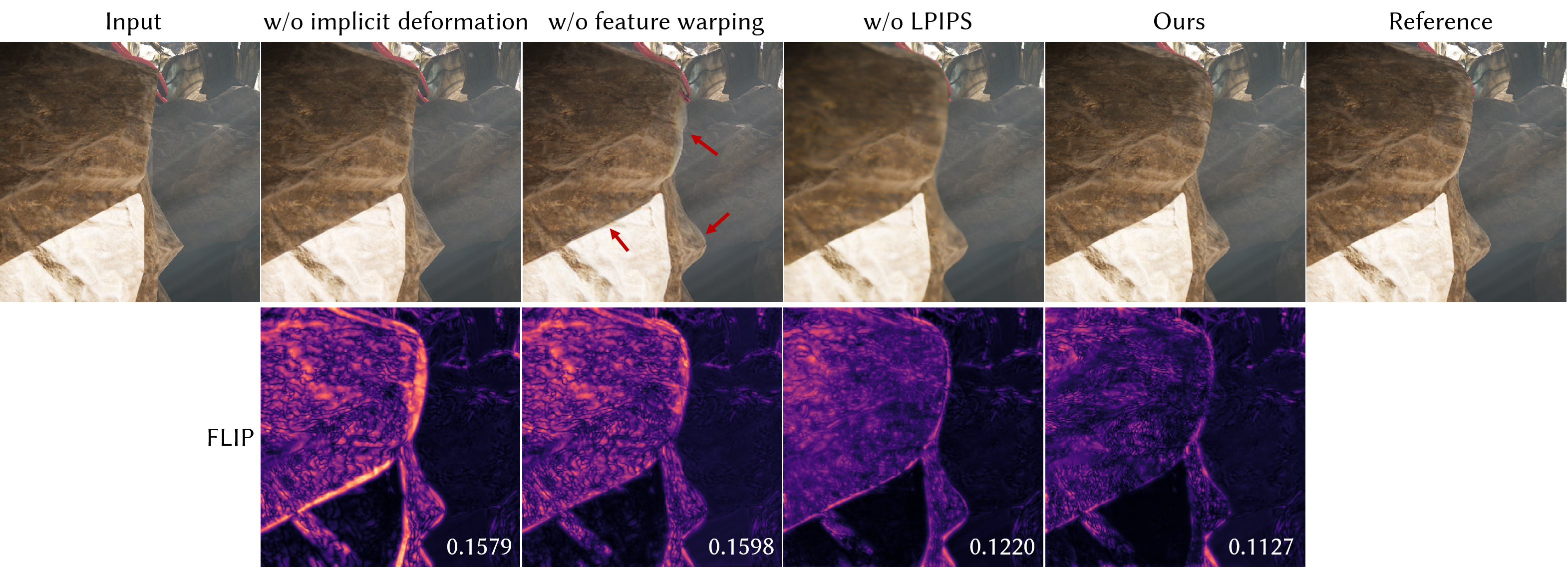}
    \caption{Ablation study of NIST. From left to right: input rendering, variants without implicit deformation, without feature warping, without LPIPS loss, our full model, and the PN-Triangles reference. Removing the implicit deformation module prevents effective silhouette refinement. Without feature warping, visible seams and appearance discontinuities emerge in newly deformed regions (highlighted by red arrows). Excluding LPIPS leads to overly smooth and blurred results. The bottom row shows FLIP error maps with respect to the reference, where lower values indicate better perceptual similarity.}
    \label{fig: ablation study}
\end{figure*}

We conduct ablation studies to evaluate the contribution of three key components in NIST: the implicit deformation module, the feature warping module, and the LPIPS perceptual loss. The results are shown in Figure \ref{fig: ablation study}.

When the implicit deformation module is removed and replaced with an equivalent lightweight $3\times3$ convolution, the network fails to learn meaningful image-space deformation, resulting in contours that remain largely unchanged from the input. Removing the feature warping module leads to visible seams between deformed and original regions, and the newly formed areas exhibit noticeable loss of appearance fidelity, which shows that a vanilla CNN structure is incapable of handling our task. Finally, when LPIPS is excluded from the training objective, the deformed regions become overly smooth and lack high-frequency details, producing visually blurred results. These ablations demonstrate that our implicit image space deformation, explicit feature warping-based reassignment, and perceptual supervision are all essential for achieving a coherent and high-fidelity NIST effect.

\subsection{Discussion and Limitations}

\paragraph{Generality.}
Our current framework is primarily trained in a per-scene manner, similar to early neural post-processing methods such as ExtraNet \cite{Guo2021ExtraNet} and DLSS 1.0 \cite{Edelsten2019TrulyNextGen}. While our current results are based on trained scene-specific models, the consistency across the evaluated scenes suggests that the learned image-space silhouette-smoothing behavior is not tied to a single scene configuration. A larger, more challenging, and more comprehensive scene dataset is predictably necessary for making this network finally become general.

\begin{figure}
    \centering
    \includegraphics[width=\linewidth]{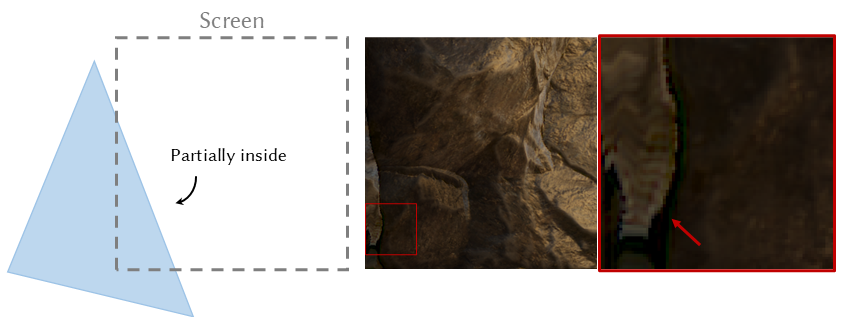}
    \caption{Screen space artifacts caused by partially visible triangles. In this case, incomplete geometric information may lead to unstable image space deformation and visible artifacts.}
    \label{fig: limitation}
\end{figure}

\paragraph{Screen-space and Temporal Limitations.}
As a screen-space method, NIST is limited by the information visible in the current frame. Partially visible triangles, thin structures, and rapidly changing screen boundaries can provide incomplete or unstable evidence for deformation, leading to local artifacts or temporal jitter near refined contours, as shown in Figure \ref{fig: limitation}. Geometry hidden from the current view cannot be inferred or deformed reliably in screen space. Future work could reduce these artifacts by incorporating temporal constraints, motion vectors, or history buffers.

\paragraph{Rendering Effects.}
NIST is currently applied to sharp rendered frames and does not explicitly model depth of field or motion blur. It also cannot correct geometry-dependent effects, such as accurate shadow silhouettes or view-dependent specular reflections, because these effects depend on geometry and visibility that are not changed by our screen-space refinement. Handling such effects would require dedicated integration into the rendering pipeline.

\paragraph{Performance.}
Our current implementation does not include industrial optimizations such as quantization. In our tested settings, the network inference cost increases modestly with image resolution and is not directly tied to the number of input mesh primitives, since it operates on rendered images and G-buffers. A complete engine-level cost analysis should also include low-poly rendering, G-buffer generation, memory transfer, and post-process integration.


\paragraph{Shading Normals.}
We use interpolated vertex normals as shading normals and do not incorporate normal maps. This reflects our focus on silhouette smoothing rather than fine-scale detail enhancement; jointly handling both remains future work.


\section{CONCLUSION AND FUTURE WORK}
We have presented Neural Image Space Tessellation effect, a lightweight screen-space post-processing approach that approximates the image-visible effect of tessellation-enabled smooth deformation without tessellating or modifying the underlying geometry. By leveraging the discrepancy between geometric face normals and interpolated vertex normals as a minimal, view-dependent cue, NIST predicts structured image-space deformation and feature warping to reduce visually apparent faceting artifacts. This design decouples the proposed silhouette refinement step from primitive count and enables its cost to scale primarily with image resolution rather than scene detail. Experimental results demonstrate that NIST produces smooth, visually coherent silhouettes close to tessellation-based smoothing references, while avoiding explicit runtime geometry refinement. In particular, our method maintains nearly constant per-frame cost across different input resolutions, with inference time remaining around 6 ms even at the highest tested resolution of 1080p.



Several directions remain open for future work. Model compression techniques such as distillation and quantization could further reduce inference, while training on larger and more diverse datasets may improve generalization across scenes and materials. In addition, incorporating explicit temporal modeling through sequence-based training could further enhance temporal stability in dynamic scenes.


\bibliographystyle{ACM-Reference-Format}
\bibliography{related}


\newpage

\end{document}